# Spin-lattice coupling in multiferroic Pb (Fe$_{1/2}$Nb$_{1/2}$)O$_3$ thin films

W. Peng[1,a)], N. Lemée[1,b)], M. Karkut[1], B. Dkhil[2], V. V. Shvartsman[3], P. Borisov[3], W. Kleemann[3], J. Holc[4], M. Kosec[4], and R. Blinc[4]

[1]Laboratoire de Physique de la Matière Condensée, Université de Picardie Jules Verne,

33 rue Saint Leu, 80039 Amiens, France,

[2]Laboratoire Structures, Propriétés et Modélisation des Solides, Ecole Centrale Paris UMR-

CNRS 8580, F-92295 Châtenary-Malabry, France,

[3]Angewandte Physik, Universität Duisburg-Essen, D-47048 Duisburg, Germany,

[4]Jožef Stefan Institute, Jamova 39, 1000 Ljubljana, Slovenia

We have made magnetization and x-ray diffraction measurements on an epitaxial Pb(Fe$_{1/2}$Nb$_{1/2}$)O$_3$ 200 nm film. From the temperature dependence of the out-of-plane lattice parameter we can assign a Burns' temperature at $T_d \sim 640$ K, a temperature at $T^* \sim 510$ K, related to the appearance of static polar nanoregions, and an anomaly occurring at 200 K. The latter is precisely the Néel temperature $T_N$ determined from magnetization and points to spin-lattice coupling at $T_N \sim 200$ K. We also observe "weak ferromagnetism" up to 300K and propose superantiferromagnetic clusters as a plausible scenario to explain this hysteresis above $T_N$.

[a)] wei.peng@u-picardie.fr

[b)]Author to whom correspondence should be address: nathalie.lemee@u-picardie.fr

Magnetoelectric (ME) multiferroics have been investigated in recent years due to the coexistence of magnetic and electric ordering parameters, exhibiting multiple functional properties. Typical ME coupling effect has been observed in various single-phase multiferroics.[1, 2] Among these, the Pb-based transition metal oxides with $ABO_3$ perovskite-type structure are of particular interest.[3] The $Pb^{2+}$ cations with a lone electron pair on the $A$-site drives off-centering displacement for ferroelectric ordering, and magnetic cations with partially filled $d$ orbitals on the $B$-site contribute to magnetic ordering. Both mechanisms are at the origin of multiferroicity.[4] Moreover, complex transition metal oxides usually exhibit fascinating cooperative electric ordering phenomena, i.e. charge, orbital and spin order, which are also believed to play significant roles in the mechanism of ME coupling.[5-7]

Lead iron niobate $Pb(Fe_{1/2}Nb_{1/2})O_3$ (PFN), a single-phase multiferroic, is a site and charge disordered relaxor ferroelectric.[8] It undergoes a paraelectric to ferroelectric phase transition at a Curie temperature $T_C \approx 385$ K[9] below which the tetragonal phase ($P4mm$) is stable down to 355K.[10] The room-temperature crystal structure has been proposed to be either monoclinic ($Cm$)[10, 11] or rhombohedral ($R3m$)[12]. There is a paramagnetic to antiferromagnetic (AF) transition at a Néel temperature $T_N \approx 145$ K.[13] To date, the observed ME effects in bulk PFN can be roughly divided into two classes. One is when the magnetic spin ordering has an effect on the dielectric properties via magnetostrictive coupling or by other electron-phonon interaction mechanisms. Some reports have confirmed the existence of a jump in the dielectric constant at $T_N$[14, 15] as well as a change in the dielectric constant induced by an external magnetic field.[16] The second is when electric dipole ordering can affect the magnetic properties due to a redistribution of the electron spins. This was demonstrated by the observation of an anomaly in the magnetic susceptibility at $T_C$.[8] Recently, taking into account the bulk lattice parameter anomalies at $T_N$ due to spin-lattice coupling,[11] a microwave dielectric spectroscopy study[17] on PFN ceramic samples revealed that the nature of the ME effect in PFN is strain mediated, i.e. it is indirectly coupling via an elastic contribution rather than by direct coupling between magnetic and electric order. Until now, most of the work relevant to ME effects in PFN has taken place on bulk samples: only a few thin film studies have been reported. Non-epitaxial PFN thin films have been fabricated by sol-gel method[18] and by pulsed laser deposition (PLD).[14] Yan *et al.*[19] recently have shown that in epitaxially grown PFN thin films there is a increase in the saturation polarization over that reported in bulk. This result gives impetus to pursuing further the epitaxial qualities of PFN and PFN-based films. In this letter, we report temperature dependent x-ray diffraction (90-800 K) and magnetization measurements (10-300 K) on a 200 nm thick epitaxial PFN film grown on a (001) $SrTiO_3$ (STO) substrate. We observe that out-of-plane lattice parameter anomalies occur at the magnetically determined $T_N$, which is direct evidence for spin-lattice coupling in PFN thin films.

The PFN films were grown using PLD with a KrF excimer laser. Details on the growth conditions have been reported elsewhere.[20] The films were analyzed by reflection high energy electron diffraction (RHEED) and by standard x-ray diffraction (XRD) using Cu $K_\alpha$ radiation. The deposition rate was determined by modeling Laue oscillations observed on thinner (< 50 nm) PFN films. Subsequently, the temperature dependence of out-of-plane lattice parameters between 90 and 800 K was investigated with an in-house high-precision diffractometer using Cu $K_\beta$ radiation. The out-of-plane lattice parameters of the films were determined from the (002) Bragg reflection after waiting for thermal equilibration at each temperature. The magnetic measurements were carried out in the temperature range of 10-300 K using a Quantum Design SQUID magnetometer.

Figure 1 shows a room temperature XRD $\theta$-$2\theta$ pattern of the PFN film. Only pseudocubic (00*l*) reflections of the film and STO substrate were observed with no detectable parasitic phases. The full width at half maximum (FWHM) for PFN (001) rocking curve is 0.09° (the STO FWHM is 0.05°), suggesting high crystalline quality with low mosaicity. In the inset of Fig.1, the $\varphi$-scan of the {220} reflection planes exhibits a fourfold symmetry, consistent with cube-on-cube epitaxial growth. This is also confirmed by RHEED streaks which were aligned with the substrate axes.[20]

In these films, phase transitions and possible lattice coupling effects are expected to induce anomalies in the out-of-plane lattice parameter. Figure 2 presents the temperature dependence of the PFN film and STO lattice parameters. The confidence error bars are less than 3% and are too small to be visible in the figure. The film is *c*-axis oriented without any splitting of the Bragg reflections over the entire temperature range. The measurements were reproducible after thermal cycling. The evolution of the lattice parameter is characterized by three anomalies at 640 K, 510 K and 200 K. We point out that there is no anomaly in STO at these characteristic temperatures, thus ruling out any substrate effect. There is no evidence for a ferroelectric phase transition around $T_C$ ≈ 385 K. In fact, the distortion reported in tetragonal PFN bulk[10] is very small ($c/a$ ≈ 1.001) and probably smaller in the film due to strain, so that it cannot be detected even with the high resolution measurements presented here. Nevertheless as evidence for ferroelectric behavior of our PFN films, local hysteresis loops were measured by piezoelectric force microscopy (not shown here). The lattice parameter anomaly at 640K, leading to a small deviation in the slope, corresponds to the Burns temperature $T_d$, at which polar nanoregions (PNRs) begin to nucleate. This relaxor-like signature is consistent with the relaxor-like behavior reported in epitaxial PFN films in Ref. 19. We point out that this result corresponds to $T_d$'s reported for other Pb-based relaxor ferroelectrics such as $Pb(Mg_{1/3}Nb_{2/3})O_3$ ($T_d$ ≈ 650 K),[21] $Pb(Fe_{2/3}W_{1/3})O_3$ ($T_d$ ≈ 650 K).[22] The second anomaly occurs around 510 K, denoted as *T\**, at which the film lattice parameters exhibits a negative thermal expansion coefficient. *T\** is a specific temperature to relaxor systems, and is related to the appearance of static PNR regions.[23] This phenomenon seems to be common in Pb-based relaxor ferroelectric systems.[24]

The third distinct lattice parameter anomaly at 200 K is, for us, the most interesting. It occurs at the magnetic phase transition temperature measured on this sample as shown in Fig. 3. On field-cooling (FC), the temperature-dependent magnetization $M$ vs $T$ curve shows a distinct jump at 200 K superimposed on a linearly increasing background, which will be discussed below. In the $dM/dT$ curve the anomaly appears as a rather pronounced minimum (inset of Fig. 3). It arises from the maximum in the susceptibility that is associated with a paramagnetic-to-AF phase transition at $T_N \sim 200$ K. For our PFN film, $T_N$ is higher than the bulk value reported at 145 K. A strain effect on $T_N$ can be ruled out, since x-ray measurements made on a thinner film (20 nm), which has a larger out-of-plane lattice parameter, also exhibits an anomaly at 200K. It has been reported[25] that the increase of high-spin $Fe^{3+}$ cation content on the $B$-site will increase the value of $T_N$. So this shift in $T_N$ in our films could be a variation in the stoichiometric ratio taking place during film growth. Since neutron[10, 12] and x-ray[11] powder diffraction studies on bulk PFN samples show that the crystal structure remains unchanged from 300 K down to 10 K, no structural phase transition is expected to contribute to the observed lattice distortion across $T_N$. In addition, the linear thermal expansion coefficient of STO is almost constant down to 150 K and thus plays no role in the lattice parameter anomaly of our film. Consequently, we conclude that the lattice parameter anomaly at and below $T_N$ is due to magnetic ordering contributions through quadratic spin-lattice coupling. This is consistent with the negative thermal expansion behavior reported in Ref. 11.

Figure 4 shows the magnetic field ($H$) dependence of the in-plane magnetization for various temperatures. The magnetization curves, $M$ vs $H$, are saturated at 300 K and at 230 K but below 200 K a linear increase of $M$ at high $H$ is observed, which indicates a contribution due to the AF ordering. However, also above $T_N$, the magnetization shows switching behavior and slim $M$-$H$ hysteresis loops are observed up to 300 K: the remanent magnetization is nonzero (inset of Fig. 4). In the absence of AF long-range order this "*weak ferromagnetism*" cannot be due to spin canting.[26] As conjectured recently[8], in the absence of AF long-range order this "*weak ferromagnetism*" is rather due to nanoparticulate magnetism since electron paramagnetic resonance (EPR) data and Langevin-type magnetization curves are in favor of fluctuating superparamagnetic (SPM) clusters of hitherto unknown origin. One possibility might be the formation of segregated ferro- or ferrimagnetic clusters incorporating mixed valences, $Fe^{3+}$ and $Fe^{2+}$, as in magnetite, $Fe_3O_4$. But this can be ruled out by the EPR data in Ref. 8 which indicate only $Fe^{3+}$ ions present in their PFN samples and for which non-zero remanent magnetization is recorded up to 340 K. Thus, owing to the inherent AF interactions within the Fe-O-Fe based magnetic subsystem of PFN, we prefer to propose a superantiferromagnetic (SAF) scenario. SAF clusters of different size following percolation statistics are assumed to gradually block on cooling and to develop weak excess moments according to Néel's theory.[27] This readily explains the observed gradual increase of the background magnetization in the $M$-$T$ curve, which continues even down to lowest temperatures (Fig. 3). The appearance of slim hysteresis then evidences SPM blocking of finite clusters with local AF rather than ferromagnetic order, but obeying conventional Stoner-Wohlfarth magnetization reversal.[28] The induced magnetization in a field of $10^4$ Oe at 10 K is ~ 30 emu/cm$^3$, and is consistent with previous work on bulk[15] and thin films.[19]

In summary, we have investigated the evolution of the lattice parameters for a 200 nm PFN film epitaxially grown on a STO substrate in the range of 90-800 K. We have determined the specific temperatures characteristic of the relaxor state: (i) $T_d$ ~ 640 K, the Burns temperature associated with the formation of polar nanoregions, (ii) $T^* \sim 510K$, the temperature related to the appearance of the static polar nanoregions. A smoothly starting negative anomaly at the Néel temperature, $T_N \sim 200$ K, coincides with observations of a related jump of the field-induced magnetization and is due to spin-lattice coupling in the PFN system. The magnetic measurements also reveal "*weak ferromagnetism*". It exists up to room temperature and is proposed to be due to blocked and switchable SAF clusters.

This work was supported by the European 6th Framework STREP: "MULTICERAL" (Grant No. FP-6-NMP-CT-2006–032616).

**FIGURE CAPTIONS**

Fig.1 XRD $\theta$-$2\theta$ pattern of a 200 nm thick PFN film on STO. In the inset a $\varphi$ scan recorded on PFN {220} reflection planes shows a cube-on-cube epitaxy.

Fig.2 (Color online) Temperature evolution of the out-of-plane lattice parameters for the PFN film (circles) and the STO substrate (diamonds). The solid lines are guides to the eyes. The characteristic temperatures for this multiferroic relaxor are indicated.

Fig.3 Temperature dependence of the magnetization of a 200 nm PFN film measured on cooling at a magnetic field of 5 kOe. The inset shows the *dM/dT* vs *T* curve.

Fig.4 (Color online) Magnetic field dependence of the magnetization at 10, 180, 230, and 300 K. The inset shows the central region of the hysteresis loops.

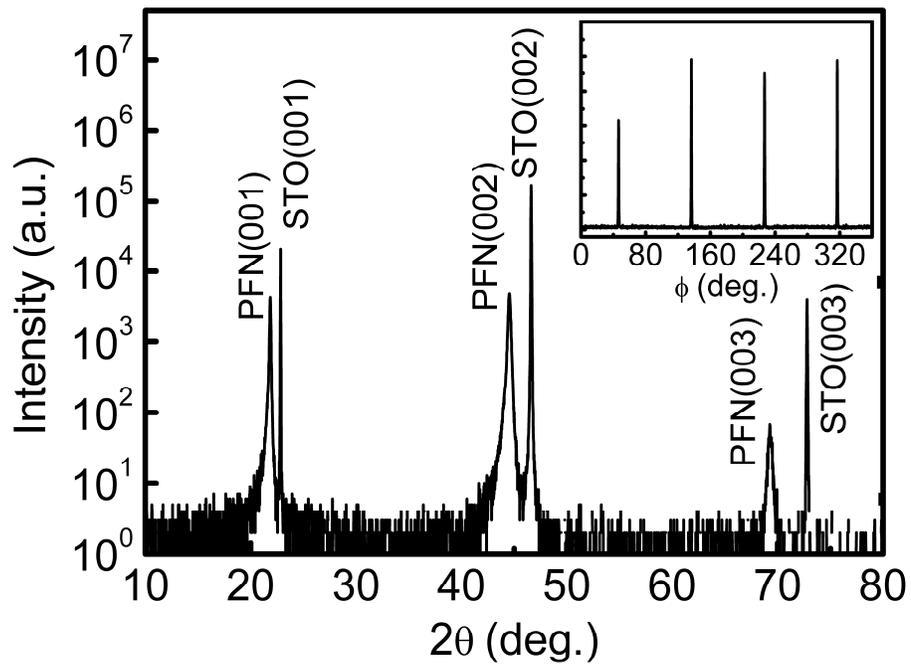

Fig.1

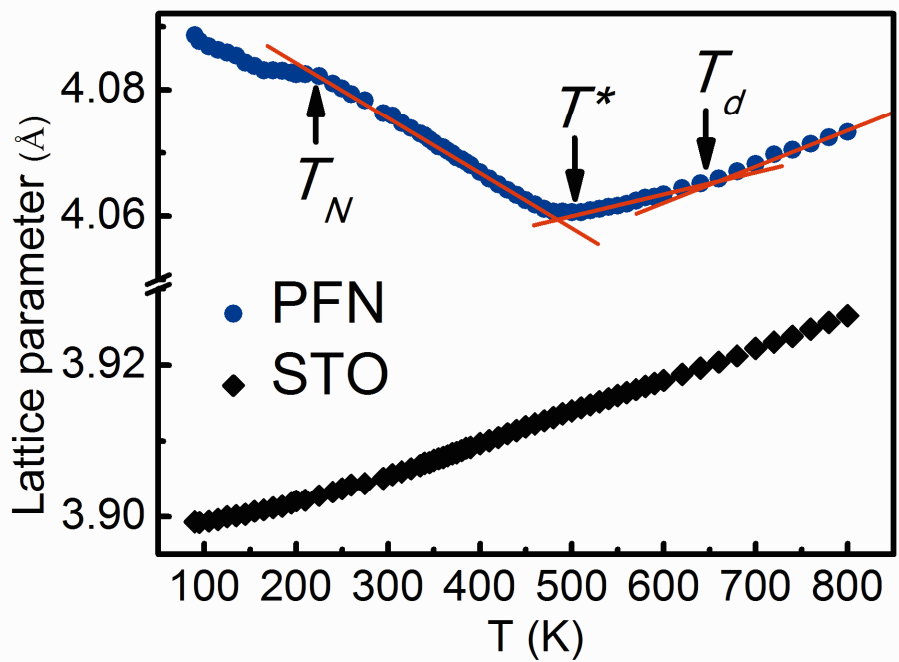

Fig.2

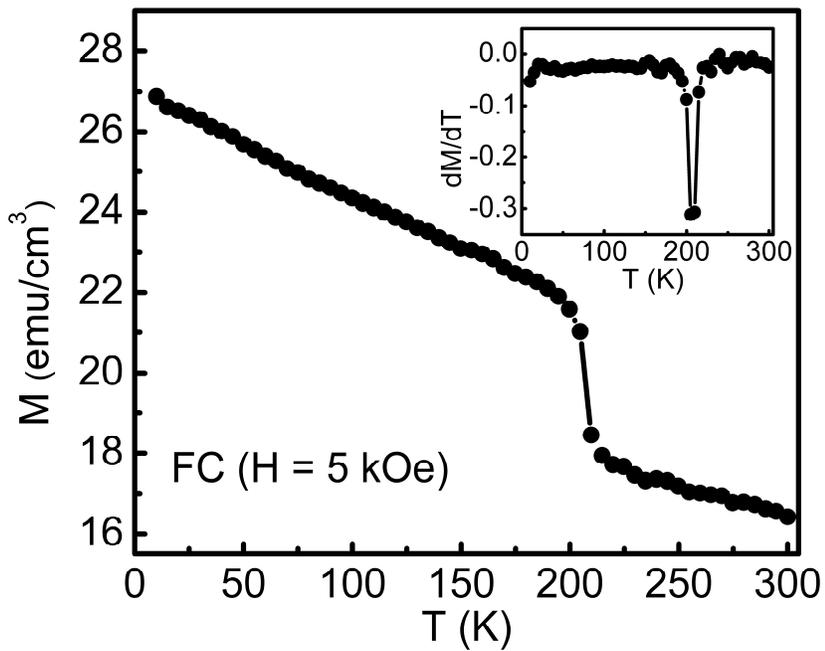

Fig.3

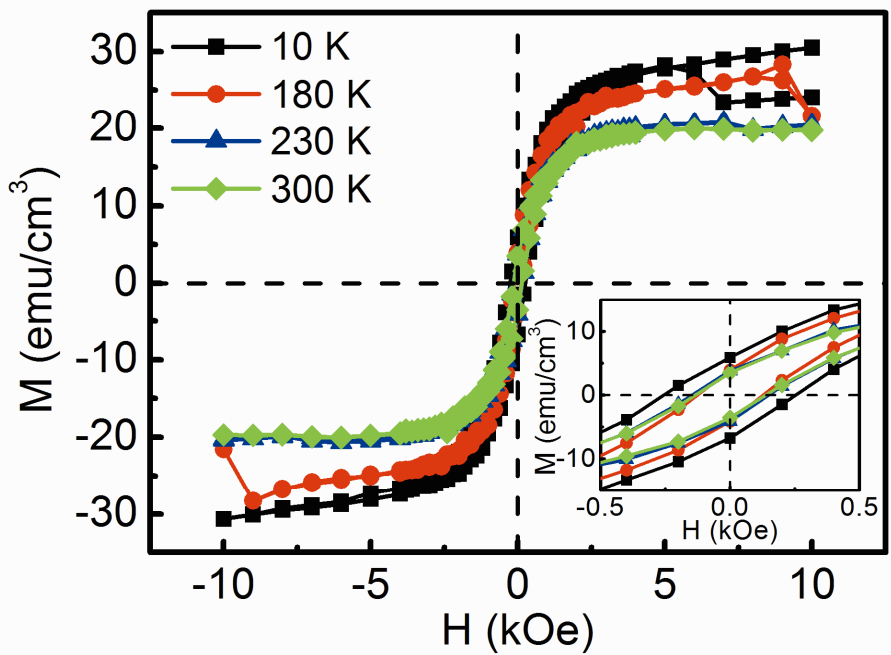

Fig.4